\documentclass[aps,prd,onecolumn,preprintnumbers,nofootinbib,10pt]{revtex4-2}
\pdfoutput=1
\usepackage{graphicx}
\usepackage{slashed}
\usepackage[dvipsnames,table]{xcolor}
\usepackage{xspace}
\usepackage[tight]{subfigure}
\usepackage{amsmath}
\usepackage{amssymb}
\usepackage{amsfonts}
\usepackage{mathrsfs}
\usepackage{comment}
\usepackage{afterpage}
\usepackage{verbatim}
\usepackage{booktabs}
\usepackage{array}
\usepackage[title,titletoc]{appendix}
\usepackage{tabularx}
\usepackage{hyperref} 

\newcolumntype{C}[1]{>{\centering\arraybackslash}p{#1}}
\def\citere#1{\mbox{Ref.~\cite{#1}}}

\newcommand{\newc}{\newcommand}
\newc{\beq}{\begin{equation}}
\newc{\eeq}{\end{equation}}
\newc{\bit}{\begin{itemize}}
\newc{\eit}{\end{itemize}}
\newc{\ben}{\begin{enumerate}}
\newc{\een}{\end{enumerate}}
\newc{\bce}{\begin{center}}
\newc{\ece}{\end{center}}
\newc{\bfi}{\begin{figure}}
\newc{\efi}{\end{figure}}

\newcommand{\ri}{\mathrm i}

\newcommand{\rd}{\mathrm d}

\newcommand{\rT}{{\mathrm{T}}}

\newcommand{\ie}{\emph{i.e.}\ }
\newcommand{\eg}{\emph{e.g.}\ }

\newcommand{\GeV}{\ensuremath{\,\text{GeV}}\xspace}
\newcommand{\TeV}{\ensuremath{\,\text{TeV}}\xspace}

\newcommand{\Pp}{\ensuremath{\text{p}}}
\newcommand{\Pe}{\ensuremath{\text{e}}\xspace}

\newcommand{\PW}{\ensuremath{\text{W}}\xspace}
\newcommand{\PZ}{\ensuremath{\text{Z}}\xspace}

\newcommand{\recola}{{\sc Recola}\xspace}

\newcommand{\PB}{\textsc{Powheg-Box}}

\newcolumntype{.}{D{.}{.}{-1}}
\newcolumntype{d}[1]{D{.}{.}{#1}}
\colorlet{tableoverheadcolor}{gray!37.5}
\colorlet{tableheadcolor}{gray!25}
\colorlet{tablerowcolor}{gray!12.5}

\def\draftdate{\relax}
\def\mda{\relax}
\def\mua{\relax}
\def\mla{\relax}
\def\draft{
\def\thtystars{******************************}
\def\sixtystars{\thtystars\thtystars}
\typeout{}
\typeout{\sixtystars**}
\typeout{* Draft mode!
         For final version remove \protect\draft\space in source file *}
\typeout{\sixtystars**}
\typeout{}
\def\draftdate{\today}
\def\mua{\marginpar[\boldmath\hfil$\uparrow$]%
                   {\boldmath$\uparrow$\hfil}\color{black}%
                    \typeout{marginpar: $\uparrow$}\ignorespaces}
\def\mda{\color{red}\marginpar[\boldmath\hfil$\downarrow$]%
                   {\boldmath$\downarrow$\hfil}%
                    \typeout{marginpar: $\downarrow$}\ignorespaces}
\def\mla{\marginpar[\boldmath\hfil$\rightarrow$]%
                   {\boldmath$\leftarrow $\hfil}%
                    \typeout{marginpar: $\leftrightarrow$}\ignorespaces}
\def\Mua{\marginpar[\boldmath\hfil$\Uparrow$]%
                   {\boldmath$\Uparrow$\hfil}\color{black}%
                    \typeout{marginpar: $\uparrow$}\ignorespaces}
\def\Mda{\color{red}\marginpar[\boldmath\hfil$\Downarrow$]%
                   {\boldmath$\Downarrow$\hfil}%
                    \typeout{marginpar: $\downarrow$}\ignorespaces}
\def\Mla{\marginpar[\boldmath\hfil\textcolor{red}{$\Rightarrow$}]%
                   {\boldmath\textcolor{red}{$\Leftarrow $}\hfil}%
                    \typeout{marginpar: $\leftrightarrow$}\ignorespaces}
\overfullrule 5pt
\oddsidemargin 15mm
\marginparwidth 29mm
}

\newcommand{\mc}{\mathcal}

\newcommand{\pt}[1]{p_{\rT,{#1}}}

\newcommand{\nnb}{\nonumber}

\begin{document}

%%%%%%%%%%%%%%%%%%%%%%%%%%%%%%%%%   
\title{SMEFT effects on spin correlations and entanglement \\at NLO QCD in di-boson production at hadron colliders} 
\author{Giovanni Pelliccioli}\email{giovanni.pelliccioli@unimib.it}   
\author{Emanuele Re}\email{emanuele.re@mib.infn.it}
\affiliation{Universit\`a degli Studi di Milano-Bicocca, Dipartimento di Fisica and\\ INFN Sezione di Milano--Bicocca, Piazza della Scienza 3, 20126 Milano, Italy} 

\begin{abstract}
We perform for the first time a full study of spin correlations in inclusive WZ production at the LHC with leptonic decays in the presence of NLO QCD corrections and of effects from a dimension-six operator in the SMEFT modifying the electroweak triple-gauge coupling. We carry out the complete quantum-state tomography of the di-boson system and relate its results to common purity and spin-entanglement markers, highlighting the sizeable impact of both QCD corrections and SMEFT insertions.
Additionally, we show how a naive truncation at dimension-six in the SMEFT expansion of the spin-density matrix can lead to a cumbersome spin interpretation of the quantum-tomography results.
\end{abstract}
\preprint{COMETA-2026-01, LAPTH-003/26}

\vspace*{0.5cm}
\maketitle
%\tableofcontents
%%%%%%%%%%%%%%%%%%%%%%%%%%%%%%%%
\section{Introduction}\label{sec:intro}

The strong interest in measuring observables inspired by quantum information (QI) at the Large Hadron Collider (LHC), further increased by the
entanglement studies by ATLAS and CMS in the top-quark sector \cite{ATLAS:2023fsd,CMS:2024pts,CMS:2024zkc}, is not limited to the
case of two-level spin states (qubits) but spans also higher dimensionalities and in particular the case of three-level spin states (qutrits).
Recently, a large number of phenomenological studies have been devoted to Higgs-boson decays
\cite{Barr:2021zcp,Aguilar-Saavedra:2022wam,Aguilar-Saavedra:2022mpg,Fabbrichesi:2023jep,Bernal:2023ruk,Fabbri:2023ncz,Bernal:2024xhm,Sullivan:2024wzl,Wu:2024ovc,Aguilar-Saavedra:2024jkj,Subba:2024aut,DelGratta:2025qyp,Aguilar-Saavedra:2025byk,Goncalves:2025xer,Aguilar-Saavedra:2025njw}
and to di-boson production at colliders 
\cite{Rahaman:2021fcz,Ashby-Pickering:2022umy,Fabbrichesi:2023cev,Aoude:2023hxv,Morales:2023gow,Subba:2024mnl,Grossi:2024jae,Grabarczyk:2024wnk,Ding:2025mzj,Goncalves:2025mvl,Ruzi:2025jql,DelGratta:2025xjp,Durupt:2025wuk,Subba:2025hxn,De:2025dpo}.
However, in spite of the availability of simulation tools at next-to-leading-order (NLO) accuracy, just a small share of works have highlighted the importance
of the inclusion of higher-order corrections in the strong (QCD) and electroweak (EW) couplings \cite{Grossi:2024jae,DelGratta:2025qyp,Aguilar-Saavedra:2025byk,Goncalves:2025mvl,Goncalves:2025xer,DelGratta:2025xjp} for an accurate evaluation of spin-correlation coefficients and QI-inspired observables that
depend on them.
Furthermore, the investigation of entanglement and Bell non-locality markers in the presence of anomalous couplings has been mostly limited to boson pairs produced in a scalar-boson decay \cite{Fabbrichesi:2023jep,Bernal:2023ruk,Sullivan:2024wzl,Subba:2024aut,DelGratta:2025qyp}.
An analysis of SMEFT effects on the entanglement markers for inclusive di-boson production at colliders has been carried in a simplified framework (no decay simulation, leading-order only) in \citere{Aoude:2023hxv}. 
Amongst di-boson production channels, $\PZ\PZ$  has attracted the greatest attention within the LHC community, owing to the high signal purity in the four-lepton decay channel, while only partial results for the quantum tomography of inclusive $\PW\PZ$ production at NLO QCD were obtained in the SM \cite{Grossi:2024jae} (see Section~(4.1) therein) and in the SMEFT \cite{Haisch:2025jqr} (see Section~(4.7) therein).
A broad and realistic study of the $\PW\PZ$ spin-density matrix and related QI-inspired observables is still missing in the literature. 
As a further motivation, the spin structure of $\PW\PZ$ production has been studied to high perturbative accuracy for what concerns single-boson and joint polarisation fractions in the so-called polarised-template method \cite{Denner:2020eck,Le:2022lrp,Pelliccioli:2025com}. On the contrary, the calculation of off-diagonal entries of the spin-density matrix for $\PW\PZ$ pairs is typically limited to the lowest order.

In this work we achieve for the first time a complete analysis of spin correlations for $\PW\PZ$ inclusive production at the LHC in the presence of NLO QCD corrections and of effects from a SMEFT dimension-six operator modifying the triple-gauge coupling. On top of the full quantum tomography, we discuss the effects of higher-order QCD corrections and SMEFT effects on relevant markers for the purity and the entanglement associated to the spin structure of the $\PW\PZ$ system.

The article is structured as follows. In Sect.~\ref{sec:analytic} we introduce the
general aspects of single- and two-boson spin-density matrices and their relation 
to the extraction of spin correlations via quantum tomography.
In Sect.~\ref{sec:SMEFT} we highlight potential issues in the interpretation of spin correlations
that arise in the presence 
of new-physics effects described by dimension-six SMEFT operators.
The numerical results are discussed in Sect.~\ref{sec:numerics},
while in Sect.~\ref{sec:conclusion} we draw our conclusions.

\section{Generalities}\label{sec:analytic}
At tree level but also including higher orders in QCD, if one assumes a two-body decay of an intermediate 
weak boson into leptons, the differential distribution in the decay angles $\theta,\phi$
has a compact analytic form \cite{Bern:2011ie,Stirling:2012zt} which can be written as an expansion in the orthonormal basis of real spherical harmonics $Y_{lm}(\Omega)$ up to rank $l=2$ \cite{Aguilar-Saavedra:2022wam}:
\begin{align}\label{eq:Vcoef}
\frac{\rd \sigma}{
\rd\Omega\,\rd {X}
}\displaystyle {\Big/}
\frac{
\rd\sigma
}{\rd X}& = 
\frac{1}{4\pi}+ \, \sum_{l=1}^2 \sum_{m=-l}^{l}\, \alpha_{lm}(X) \, Y_{lm}\big (\Omega\big ) \,,\qquad \rd\Omega=\rd\phi\,\rd\!\cos\!\theta\,,
\end{align} 
where $X$ is a generic observable that is independent of the decay angular measure, \eg the transverse momentum of the decaying boson. 
Since we are interested in di-boson systems, we define decay angles in the so-called \emph{modified helicity coordinate system} \cite{Aaboud:2019gxl}, \ie $\theta,\phi$ are the polar and azimuthal angles of the positively charged lepton in the corresponding boson rest frame,
computed taking as reference axis for the polar angle $\theta$ the boson spatial direction in the di-boson centre-of-mass (CM) frame.
In the absence of kinematic constraints on the decay products, the $\alpha_{lm}$ coefficients can be extracted by projecting onto corresponding spherical harmonics,\\[-0.4cm]
\begin{align}\label{eq:VcoefPROJ}
\alpha_{lm}(X) = \cfrac{
\displaystyle\int\!
\rd \Omega\, \cfrac{\rd \sigma}{
\rd\Omega\,\rd X}\,Y_{lm}(\Omega)
}{
\displaystyle\int\! \rd \Omega\, \cfrac{\rd \sigma}{
\rd\Omega\,\rd X}
}&  \,.
\end{align}
To relate the angular coefficients to the spin-density matrix for a single boson $V$ we rewrite Eq.~\eqref{eq:Vcoef} as,
\beq\label{eq:RhoGamma}
\frac{\rd \sigma}{
\rd\Omega\,\rd {X}
}\displaystyle {\Big/}
\frac{
\rd\sigma
}{\rd X}
\,=\,\frac3{4\pi}
\,\sum_{i,j=1}^3\rho_{V,ij}(X)
\Gamma_{ij}(\Omega)
\,,
\eeq
where $\Gamma$ is the decay matrix for the two-body decay to massless leptons \cite{Boudjema:2009fz,Rahaman:2021fcz}, 
\beq\label{eq:gamma}
\displaystyle\Gamma(\Omega)\,=\,\frac14\left(\begin{array}{ccc}
     {1+2\eta_\ell\cos\theta+\cos^2\theta} & \sqrt{2}\,\sin\theta(\eta_\ell + \cos\theta)e^{\ri \phi}  & (1-\cos^2\theta) e^{2\ri \phi}\\
     \sqrt{2}\,\sin\theta(\eta_\ell + \cos\theta) e^{-\ri \phi}
     & 2\,({1-\cos^2\theta}) & \sqrt{2}\,\sin\theta(\eta_\ell - \cos\theta) e^{\ri \phi}\\
     (1-\cos^2\theta) e^{-2\ri \phi}& 
     \sqrt{2}\,\sin\theta(\eta_\ell - \cos\theta)e^{-\ri \phi}& {1-2\eta_\ell\cos\theta+\cos^2\theta}
\end{array}\right)\,,
\eeq
and $\rho_V$ is a linear combination of the identity $\mathbb{I}_3$ and of the irreducible tensor representations for qutrits $T_{lm}$ \cite{Aguilar-Saavedra:2022wam,DelGratta:2025qyp}, \footnote{The $T_{lm}$ matrices are defined such that $T_{l,-m}=(-1)^m\,T^{\dagger}_{lm}$, and read,
\begin{equation}
T_{11} = \sqrt{\frac{3}{2}}
\begin{pmatrix}
0 & -1 & 0 \\
0 & 0 & -1 \\
0 & 0 & 0
\end{pmatrix}, \quad
T_{10} = \sqrt{\frac{3}{2}}
\begin{pmatrix}
1 & 0 & 0 \\
0 & 0 & 0 \\
0 & 0 & -1
\end{pmatrix},%\nnb
\quad
T_{22} = \sqrt{3}
\begin{pmatrix}
0 & 0 & 1 \\
0 & 0 & 0 \\
0 & 0 & 0
\end{pmatrix}, \quad
T_{21} = \sqrt{\frac{3}{2}}
\begin{pmatrix}
0 & -1 & 0 \\
0 & 0 & 1 \\
0 & 0 & 0
\end{pmatrix}, \quad
T_{20} = \frac{1}{\sqrt{2}} \begin{pmatrix}
1 & 0 & 0 \\
0 & -2 & 0 \\
0 & 0 & 1
\end{pmatrix}.\nnb
\end{equation}
}
\begin{equation}
\rho_V(X) = \frac13 \left[\mathbb{I}_3  + \displaystyle\sum_{l=1}^2\kappa_{l}\sum_{m=-l}^l \alpha_{lm}(X) \,T_{lm}  \right]\,,\qquad \kappa_1 = \frac{\sqrt{8\pi}}{\eta_\ell},\quad\kappa_2=\sqrt{40\pi}\,.
\label{eq:rhoexp}
\end{equation}
We note that the potential dependence on the kinematics of the decaying boson ($X$) enters the production spin-density matrix $\rho_V$ but not the decay matrix $\Gamma$.  However, it was recently shown \cite{DelGratta:2025xjp} that a mild dependence on the boson transverse momentum can be present in the spin-analysing power $\eta_\ell$ in Eq.~\ref{eq:gamma}, owing to lepton dressing starting at NLO EW. Since in this work we only consider QCD corrections we can safely assume that $\Gamma$ is completely independent of any $X$ other than the decay angles. Furthermore, the analytic structure of Eq.~\ref{eq:gamma} is valid in the presence of higher-order QCD corrections, and to large extent also of radiative EW corrections, up the modification of the numerical value of the spin-analysing power $\eta_\ell$ \cite{DelGratta:2025xjp}.

The spin-density matrix in Eq.~\eqref{eq:rhoexp} (but also its generalisation to a system of two or more bosons) is hermitian, has unit trace, and is semi-positive definite. Being in general a mixture of pure-state projectors, it can be written in terms of helicity amplitudes \cite{Barr:2024djo} as follows,
\beq\label{eq:rhoSM}
\rho_{V,ij} = 
\frac{
\phantom{\Big(}\displaystyle \sum_x\mc A_{i,x}\mc A^{*}_{j,x}\phantom{\Big)}
}{\displaystyle \sum_x\sum_{a=1}^3 \left|\mc A^{}_{a,x}\right|^2}
\,,\qquad
{\rm Tr}\left[{\rho_V}\right]= 1
\,,\qquad
\frac13 \leq {\rm Tr}\left[{\rho_V}^2\right]\leq 1\,,
\eeq  
where the sum over $x$ understands sums and averages over all degrees of freedom other than the helicity of the considered boson.
We note that ${\rm Tr}\left[{\rho_V}^2\right] = 1$ is achieved for pure states, while $ {\rm Tr}\left[{\rho_V}^2\right]\geq \frac13 $ is a direct consequence of the Cauchy-Schwartz inequality.
From Eq.~\eqref{eq:rhoexp}, we obtain the following expression for the trace of $\rho_V^2$ in terms of $\alpha_{lm}$ coefficients (making the dependence on independent variables $X$ implicit):
\beq\label{eq:trV}
{\rm Tr}\left[{\rho_V}^2\right]=\frac13\left[
1 + \displaystyle\sum_{l=1}^2\sum_{m=-l}^l \left(\kappa_{l}\alpha_{lm}\right)^2
\right]\,,
\eeq
which can be used to assess the purity of generic mixed states.

The generalisation of Eq.~\eqref{eq:Vcoef} to the case of two EW bosons reads \cite{Aguilar-Saavedra:2022wam,Grossi:2024jae,DelGratta:2025qyp},
\begin{align}\label{eq:VVcoef}
\frac{\rd \sigma}{
\rd\Omega_1\,\rd\Omega_2\,\rd {X}
}\displaystyle {\Big/}
\frac{
\rd\sigma
}{\rd X}& = 
\frac{1}{(4\pi)^2}
+ 
\, \frac{1}{4\pi}\sum_{l=1}^2 \sum_{m=-l}^{l}\, \alpha^{(1)}_{lm}(X) \, Y_{lm}\big (\Omega_1\big )
+\, \frac{1}{4\pi}\sum_{l'=1}^2 \sum_{m='-l'}^{l'}\, \alpha^{(2)}_{l'm'}(X) \, Y_{l'm'}\big (\Omega_2\big )
\nnb\\
&+\, \sum_{l,l'=1}^2 \sum_{m=-l}^{l} \sum_{m'=-l'}^{l'}\, \gamma_{lml'm'}(X) \, Y_{lm}\big (\Omega_1\big )Y_{l'm'}\big (\Omega_2\big )
\,,
\end{align} 
where $\Omega_1,\Omega_2$ represent the decay angles associated to the first and second boson, respectively. 
All polarisation ($\alpha^{(1)}_{lm}, \alpha^{(2)}_{lm}$) and spin-correlation ($\gamma_{lml'm'}$) coefficients can 
be extracted from Eq.~\eqref{eq:VVcoef} by projecting onto one or two spherical harmonics, in the same fashion as in Eq.~\eqref{eq:VcoefPROJ}.
Following the same reasoning as above, the four-dimensional angular distribution in Eq.~\eqref{eq:VVcoef}
can also be written as a combination of the spin-density matrix $\rho$ for two qutrits and two decay matrices, extending Eq.~\eqref{eq:RhoGamma},
\beq\label{eq:VVtrace}
\frac{\rd\sigma}{\rd\Omega_1\,\rd\Omega_2\,\rd X} \displaystyle {\Big/}\frac{\rd\sigma}{\rd X}\,=\,
\left(\frac{3}{4\pi}\right)^2\,\sum_{i,i'\!,j,j'=1}^3\rho_{i\,i'jj'}(X)
\Gamma_{ii'}(\Omega_1)
\Gamma_{jj'}(\Omega_2)\,,
\eeq
where the decay matrices have the form of Eq.~\eqref{eq:gamma} and $\rho$ is now a hermitian, unit-trace, 9$\times$9 matrix that reads,
\begin{align}
\rho(X) &= \frac19 \bigg[\mathbb{I}_3 \otimes \mathbb{I}_3  
+ \displaystyle\sum_{l=1}^2\kappa^{(1)}_{l}\sum_{m=-l}^l \alpha^{(1)}_{lm}(X) \,\,T_{lm} \otimes  \mathbb{I}_3 \,
+ \displaystyle\sum_{l'=1}^2\kappa^{(2)}_{l'}\sum_{m'=-l'}^{l'} \alpha^{(2)}_{l'm'}(X) \,\,\mathbb{I}_3 \otimes T_{l'm'}  \,
\nnb\\
&\hspace{1cm}+ \displaystyle\sum_{l,l'=1}^2\kappa^{(1)}_{l}\kappa^{(2)}_{l'}\sum_{m=-l}^l\sum_{m'=-l'}^{l'} \gamma_{lml'm'}(X) \,\,T_{lm} \otimes T_{l'm'}  \bigg]\,,
\qquad \kappa^{(i)}_1 = \frac{\sqrt{8\pi}}{\eta_{\ell,i}},\quad\kappa^{(i)}_2=\sqrt{40\pi}\,.
\label{eq:rhoexpVV}
\end{align}
Note that for $V_1=\PW,\,V_2=\PZ$ the two spin-analysing powers are different, $\eta_{\ell,\PW}\neq\eta_{\ell,\PZ}$.
Although we do not write it explicitly, the $\rho$ matrix for two bosons can be easily expressed in terms of helicity amplitudes similarly to Eq.~\eqref{eq:rhoSM} by making two helicity indices explicit (one for each boson).
The purity of the di-boson state can be assessed by taking the trace of the squared of the $\rho$ matrix, which reads \cite{DelGratta:2025qyp},
\beq\label{eq:tr2VV}
{\rm Tr}\left[{\rho}^2\right]=\frac19\left[
1 + \displaystyle\sum_{l=1}^2\sum_{m=-l}^l \left(\kappa^{(1)}_{l}{\alpha^{(1)}_{lm}}\right)^2
+ \displaystyle\sum_{l'=1}^2\sum_{m'=-l'}^{l'} \left(\kappa^{(2)}_{l'}{\alpha^{(2)}_{l'm'}}\right)^2
+ \displaystyle\sum_{l,l'=1}^2\sum_{m=-l}^l\sum_{m'=-l'}^{l'}  \left(\kappa^{(1)}_{l}\kappa^{(2)}_{l'}{\gamma_{lml'm'}}\right)^2
\right]\,.
\eeq

From Eqs.~\eqref{eq:rhoexp} and \eqref{eq:rhoexpVV} it is clear that by extracting the coefficients with $\ell \leq 2$ from angular distributions in Eq.~\eqref{eq:Vcoef} and Eq.~\eqref{eq:VVcoef}, one can entirely reconstruct the structure of the spin-density matrix for one ($\rho_V$) and two ($\rho$) EW bosons, or qutrits. This approach goes under the name of quantum-state tomography (QT).
From the full knowledge of the spin-density matrix one can construct suitable markers for entanglement, Bell non-locality, and other QI-inspired observables. In this work we focus on spin entanglement.
The level of entanglement for two-qutrit systems can be evaluated through the concurrence \cite{Hill:1997pfa,Mintert:2007pow,Zhang:2008ebu,Horodecki:2009zz}
for which a closed analytic expression only exists in special configurations. 
In the case of generic mixed states, the concurrence $\mc C$ is defined through an optimisation procedure and only lower and upper bounds are known \cite{Mintert:2007pow,Zhang:2008ebu}. These can be expressed as \cite{DelGratta:2025qyp},
\beq\label{eq:concurrence}
2\,{\rm max}\Big({\rm Tr}[\rho^2]-{\rm Tr}[\rho_{V_1}^2],{\rm Tr}[\rho^2]-{\rm Tr}[\rho_{V_2}^2]\Big)
\leq
{\mc C}^2 
\leq
2\,{\rm min}\Big(1-{\rm Tr}[\rho_{V_1}^2],1-{\rm Tr}[\rho_{V_2}^2]\Big)\,,
\eeq
where the trace of squared spin-density matrices for individual bosons and for the boson pair take the form of Eq.~\eqref{eq:trV} and Eq.~\eqref{eq:tr2VV}, respectively.
If the lower bound in Eq.~\eqref{eq:concurrence} is larger than zero, then the two-boson spin state is entangled.

\section{SMEFT effects} \label{sec:SMEFT}
The whole treatment above applies 
as long as the probabilistic interpretation of the spin-density matrix (semi-positivity) is not jeopardised. This holds independently of the specific dynamical theory underlying the EW-boson production, whether the SM or any beyond-the-SM theory.

We now focus on potential new-physics effects modeled in the framework of the Standard-Model Effective Field Theory (SMEFT).
For simplicity, but without loss of generality of our conclusions, we consider one of the least constrained CP-even operators leading to anomalous triple-gauge couplings (aTGC) in the EW sector and often studied in the context of di-boson processes \cite{Falkowski:2015jaa,Falkowski:2016cxu,Azatov:2016sqh,Helset:2017mlf,Azatov:2017kzw,Baglio:2017bfe,Panico:2017frx,Chiesa:2018lcs,Baglio:2018bkm,Grojean:2018dqj,Baglio:2019uty,Azatov:2019xxn,Baglio:2020oqu,Degrande:2024bmd,ElFaham:2024uop,Haisch:2025jqr},
\begin{equation}\label{eq:QW}
{\mc L}_{\rm SMEFT}= {\mc L}_{\rm SM}+ \frac{C_{W}}{\Lambda^2}\hspace{0.5mm}Q_W \,,\qquad
    Q_{W} = \epsilon_{ijk} \hspace{0.25mm} W^{i, \nu}_{\mu} \hspace{0.25mm} W^{j, \lambda}_{\nu} \hspace{0.25mm}W^{k, \mu}_{\lambda}\,,
\end{equation}
where $C_W$ is the Wilson coefficient (WC) associated to the $Q_W$ operator and $\Lambda$ is the scale of new physics. 
Up to $\Lambda^{-4}$, and with a single insertion of the considered operator, a standard observable, \eg the transverse momentum of the EW boson, would receive contributions from squared SM ($\Lambda^0$) amplitudes, from the linear interference between SM and EFT amplitudes ($\Lambda^{-2}$), and from squared EFT amplitudes ($\Lambda^{-4}$),
\begin{equation}
\cfrac{\rd \sigma}{\rd X}=
\cfrac{\rd \sigma^{(\rm 4)}}{\rd X}
+\xi\,\cfrac{\rd \sigma^{(6)}}{\rd X}
+\xi^2\,\cfrac{\rd \sigma^{(8)}}{\rd X}\,,
\end{equation}
where we define $\xi = C_W/\Lambda^2$ and the canonical dimensions of the three contributions are used in the labels ($(4)$ for the SM, $(6)$ for the linear SM--EFT interference, and $(8)$ for the quadratic EFT term). The impact on the angular coefficients is however more intricate, as the dependence on the SMEFT effects enters both in the numerator and in the denominator of Eq.~\eqref{eq:VcoefPROJ} and therefore they are extracted as \cite{ElFaham:2024uop,ElFaham:2025fow},
\begin{align}\label{eq:VcoefPROJeft}
\alpha_{lm}(X) = \cfrac{
\displaystyle\int\!
\rd \Omega\, 
\left[
\cfrac{\rd \sigma^{(\rm 4)}}{\rd\Omega\,\rd X}
+\xi\,\cfrac{\rd \sigma^{(6)}}{\rd\Omega\,\rd X}
+\xi^2\,\cfrac{\rd \sigma^{(8)}}{\rd\Omega\,\rd X}
\right]
\,Y_{lm}(\Omega)
}{
\displaystyle\int\! \rd \Omega\, 
\left[
\cfrac{\rd \sigma^{(\rm 4)}}{\rd\Omega\,\rd X}
+\xi\,\cfrac{\rd \sigma^{(6)}}{\rd\Omega\,\rd X}
+\xi^2\,\cfrac{\rd \sigma^{(8)}}{\rd\Omega\,\rd X}
\right]
}&  \,.
\end{align} 
\begin{comment}
\begin{align}\label{eq:VVcoef}
\frac1 {\sigma}\frac{\rd \sigma}{
\rd\!\cos\theta_{1}
\,\rd\phi_{1}
\,\,\rd\!\cos\theta_{2}
\,\rd \phi_{2}
}  = 
\frac{1}{(4\pi)^2}
&+ \frac{1}{4\pi}\, \sum_{l=1}^2 \sum_{m=-l}^{l}\, \alpha^{(1)}_{lm} \, Y_{lm}\big (\theta_{1}, \phi_{1}\big ) + \frac{1}{4\pi}\, \sum_{l=1}^2 \sum_{m=-l}^{l}\, \alpha^{(2)}_{lm} \, Y_{lm}\big (\theta_{2},\phi_{2}\big )
\nonumber \\[2mm]
& + \sum_{l=1}^2 \sum_{l^\prime=1}^2 \sum_{m=-l}^{l}\sum_{m^\prime=-l^\prime}^{l^\prime}\, \gamma_{lm l^\prime m^\prime} \, Y_{lm}\big (\theta_{1}, \phi_{1}\big ) \, Y_{l^\prime m^\prime}\big (\theta_{2}, \phi_{2}\big ) \,,
\end{align} 
\end{comment}
In the SMEFT the generic amplitude $\mc A^{}_i$ reads $\mc A^{(4)}_i+\xi\mc A^{(6)}_i$, therefore the spin-density matrix in Eq.~\eqref{eq:rhoSM} becomes,
\beq\label{eq:rhoSMEFT8}
\rho^{(\rm 8)}_{V,ij} = \frac{ 
\displaystyle\sum_{x}
\left(\mc A^{(4)}_{i,x}+\xi \mc A^{(6)}_{i,x}\right)
\left(\mc A^{(4)*}_{j,x}+\xi \mc A^{(6)*}_{j,x}\right)
}{
\displaystyle\displaystyle\sum_{x}\sum_{a=1}^3 
\left|\mc A^{(4)}_{a,x} + \xi \mc A^{(6)}_{a,x}\right|^2
}\,=\,
\frac{ 
\displaystyle\sum_{x}
\left[
\left(\mc A^{(4)}_{i,x}\mc A^{(4)*}_{j,x}\right)
+\xi
\left(\mc A^{(4)}_{i,x}\mc A^{(6)*}_{j,x}+\mc A^{(6)}_{i,x} \mc A^{(4)*}_{j,x}\right)
+ 
\xi^2\left(\mc A^{(6)}_{i,x}\mc A^{(6)*}_{j,x}\right)\right]
}{
\displaystyle\sum_{x}\displaystyle\sum_{a=1}^3 \left[
\left|\mc A^{(4)}_{a,x}\right|^2
+2 \,\xi \,{\mathfrak{Re}}\!\left(\mc A^{(4)}_{a,x} \mc A^{(6)*}_{a,x}\right)
+\xi^2\left|\mc A^{(6)}_{a,x}\right|^2
\right]
}\,,
\eeq
where all terms in the SMEFT $\xi$ expansion are retained, including quadratic ones. Just like Eq.~\eqref{eq:rhoSM}, the SMEFT spin-density matrix in Eq.~\eqref{eq:rhoSMEFT8} is hermitian, has unit trace, and is semi-positive definite.

The SMEFT expansion is often studied up to linear order in $\xi$, \ie excluding any effect which is formally beyond order $\Lambda^{-2}$. This requires to ignore quadratic terms from Eq.~\eqref{eq:VcoefPROJeft}, leading to,
\begin{align}\label{eq:VcoefPROJeft6}
\alpha_{lm}(X) = \cfrac{
\displaystyle\int\!
\rd \Omega\, 
\left[
\cfrac{\rd \sigma^{(\rm 4)}}{\rd\Omega\,\rd X}
+\xi\,\cfrac{\rd \sigma^{(6)}}{\rd\Omega\,\rd X}
\right]
\,Y_{lm}(\Omega)
}{
\displaystyle\int\! \rd \Omega\, 
\left[
\cfrac{\rd \sigma^{(\rm 4)}}{\rd\Omega\,\rd X}
+\xi\,\cfrac{\rd \sigma^{(6)}}{\rd\Omega\,\rd X}
\right]
}&  \,,
\end{align} 
If the same is done for the spin-density matrix, the quadratic contributions in $\xi$ have to be discarded both in the numerator and in the denominator of Eq.~\eqref{eq:rhoSMEFT8}, leading to,
\beq\label{eq:rhoSMEFT6}
\rho^{(\rm 6)}_{V,ij} \,=\,
\frac{ 
\displaystyle\sum_{x}
\left[
\left(\mc A^{(4)}_{i,x}\mc A^{(4)*}_{j,x}\right)
+\xi
\left(\mc A^{(4)}_{i,x}\mc A^{(6)*}_{j,x}+\mc A^{(6)}_{i,x} \mc A^{(4)*}_{j,x}\right)\right]
}{
\displaystyle\sum_{x}\displaystyle\sum_{a=1}^3 \left[
\left|\mc A^{(4)}_{a,x}\right|^2
+2 \,\xi \,{\mathfrak{Re}}\!\left(\mc A^{(4)}_{a,x} \mc A^{(6)*}_{a,x}\right)
\right]
}\,.
\eeq
While the $\rho_V$ matrix in Eq.~\eqref{eq:rhoSMEFT6} is hermitian and has unit trace, it is not anymore semi-positive definite, leading to important consequences in the interpretation of the angular analysis.

To show that the spin-density matrix at dimension-six in SMEFT like in Eq.~\eqref{eq:rhoSMEFT6} is ill-defined, we tailor the discussion to the simplified case of a pure state, for which there is no sum over additional degrees of freedom $x$. 
It is straightforward to show that the $\rho_V^{(8)}$ matrix for a pure state,
\beq\label{eq:temp}
\rho^{(\rm 8,\,pure)}_{V,ij} = \frac{ 
\left(\mc A^{(4)}_{i}+\xi \mc A^{(6)}_{i}\right)
\left(\mc A^{(4)*}_{j}+\xi \mc A^{(6)*}_{j}\right)
}{
\displaystyle\sum_{a=1}^3 
\left|\mc A^{(4)}_{a} + \xi \mc A^{(6)}_{a}\right|^2
}\,=\,
\frac{ 
\left(\mc A^{(4)}_{i}\mc A^{(4)*}_{j}\right)
+\xi
\left(\mc A^{(4)}_{i}\mc A^{(6)*}_{j}+\mc A^{(6)}_{i} \mc A^{(4)*}_{j}\right)
+ 
\xi^2\left(\mc A^{(6)}_{i}\mc A^{(6)*}_{j}\right)
}{
\displaystyle\sum_{a=1}^3 \left(
\left|\mc A^{(4)}_{a}\right|^2
+2 \,\xi \,{\mathfrak{Re}}\!\left(\mc A^{(4)}_{a} \mc A^{(6)*}_{a}\right)
+\xi^2\left|\mc A^{(6)}_{a}\right|^2
\right)
}\,.
\eeq
is a projector (${\rm Tr}[\rho_V^2]=1$) for any value of $\xi$.
If we take Eq.~\ref{eq:rhoSMEFT6} and tailor it to a pure state, we get,
\beq\label{eq:temp2}
\rho^{(\rm 6,\,pure)}_{V,ij} \,=\,
\frac{ 
\left(\mc A^{(4)}_{i}\mc A^{(4)*}_{j}\right)
+\xi
\left(\mc A^{(4)}_{i}\mc A^{(6)*}_{j}+\mc A^{(6)}_{i} \mc A^{(4)*}_{j}\right)
}{
\displaystyle\sum_{a=1}^3 \left(
\left|\mc A^{(4)}_{a}\right|^2
+2 \,\xi \,{\mathfrak{Re}}\!\left(\mc A^{(4)}_{a} \mc A^{(6)*}_{a}\right)
\right)
}\,,
\eeq
then the trace of its squared reads,
\begin{align}
{\rm Tr}[{\rho_V^{\rm (6\,,pure)}}^2] 
&=
\frac{ 
\displaystyle\sum_{i,j=1}^3\Bigg|
\left(\mc A_{i}\mc A^{*}_{j}\right)
+
\left(\mc A_{i}\mc F^{*}_{j}+\mc F_{i} \mc A^{*}_{j}\right)
\Bigg|^2
}{
\left[\displaystyle\sum_{a=1}^3 \left(
\left|\mc A_{a}\right|^2
+\, \left(\mc A_{a} \mc F^{*}_{a}+\mc A^{*}_{a} \mc F_{a}\right)
\right)
\right]^2}
\nnb\\
&= \frac{\displaystyle\sum_{i,j=1}^3
\bigg[
|\mc A_i|^2|\mc A_j|^2
+
2|\mc A_i|^2 (\mc A_j \mc F_j^* + \mc A_j^* \mc F_j)
+
2|\mc A_i|^2|\mc F_j|^2
+
\mc A_i \mc A_j \mc F_i^* \mc F_j^*
+
\mc A_i^* \mc A_j^* \mc F_i \mc F_j
\bigg]}{
\displaystyle\sum_{i,j=1}^3 \left[
\left|\mc A_{i}\right|^2 \left|\mc A_{j}\right|^2
+
2\left|\mc A_{i}\right|^2 (\mc A_{j} \mc F^{*}_{j}+\mc A^{*}_{j}\mc F_{j})
+
2\mc A_{i} \mc A^{*}_{j} \mc F^{*}_{i}\mc F_{j}
+
\mc A_{i}\mc A_{j}  \mc F^{*}_{i}\mc F^{*}_{j}
+
\mc A^{*}_{i}\mc A^{*}_{j}  \mc F_{i}\mc F_{j}
\right ]\label{eq:proofTR6}
} 
\end{align}
where we have renamed $\mc A^{(4)}\rightarrow \mc A$ and $\xi \mc A^{(6)}\rightarrow \mc F$.
Owing to the Cauchy-Schwartz inequality, 
\beq
\left(\displaystyle\sum_{i=1}^3 |\mc A_i|^2\right)\left(\displaystyle\sum_{i=1}^3|\mc F_j|^2\right)
\quad \geq \quad
\left|\displaystyle\sum_{i=1}^3 \mc A_{i}\mc F^{*}_{i} \right|^2\,,
\eeq
in Eq.~\eqref{eq:proofTR6} the numerator can exceed or equal the denominator, therefore,
\beq
{\rm Tr}[{\rho_V^{\rm (6\,,pure)}}^2] \geq 1\,.
\eeq
This implies that, independently of the helicity rules in the SM and in the EFT amplitudes, the truncation of the density-matrix SMEFT expansion at linear level does not allow for a straightforward interpretation as a suitable spin-density matrix.
Since such an ill definition of the spin-density matrix holds for a pure state, it cannot be excluded that the same holds for mixed states as well (see Eq.~\eqref{eq:rhoSMEFT6}), especially in kinematic regions
and dynamical regimes %, 
where the purity of the state is high, as we will see in the next section.
On the contrary, the spin-density definitions in Eqs.~\eqref{eq:rhoSMEFT8} and \eqref{eq:temp} correctly satisfies
\beq
{\rm Tr}\left[  {\rho^{(\rm 8)}_{V,ij} }^2\right] \leq 1
\,,\qquad 
{\rm Tr}\left[  {\rho^{(\rm 8,\,pure)}_{V,ij} }^2\right] = 1\,,
\eeq
in arbitrary kinematic and dynamical regimes.
As a natural consequence, also the spin-density matrix for a di-boson (two-qutrit) state is affected by similar issues, when truncating the SMEFT expansion at linear level in the WC of SMEFT operators. 
In such case, as for a single boson (see Eq.~\eqref{eq:rhoSMEFT6}), the interpretation of the angular coefficients as spin correlations of the di-boson system becomes cumbersome. In particular, we will see
from the numerical results that the trace in Eq.~\eqref{eq:tr2VV} can exceed the value of one for certain kinematical regimes and certain value of the WC under consideration.

\section{Numerical results}\label{sec:numerics}
We consider the inclusive production of $\PW\PZ$ pairs in Run-3 LHC proton--proton collisions ($\sqrt{s}=13.6~\TeV$).
Although it was shown \cite{Grossi:2024jae,Goncalves:2025mvl,DelGratta:2025xjp} that the full off-shell picture of angular coefficients in di-boson processes is correctly reproduced by factorised calculations with intermediate on-shell bosons, we perform all calculations in the double-pole approximation with the newly released di-boson package \cite{Chiesa:2020ttl,Pelliccioli:2023zpd,Haisch:2025jqr} in the \PB~ framework \cite{Nason:2004rx,Frixione:2007vw,Alioli:2010xd,Jezo:2015aia}, ensuring that intermediate spin-1 bosons are on shell.
Specifically we consider the decay channel with three massless charged leptons and one neutrino,
\beq\label{eq:WZproc}
\Pp\Pp \to \PW^+(\to \Pe^+\nu_{\Pe})\,\PZ( \to\mu^+\mu^-)\,.
\eeq
The calculation is carried at both LO and at NLO QCD perturbative accuracy,
while parton-shower effects are not considered, since it was proved that they have a mild impact on the angular coefficients \cite{Haisch:2025jqr}.
We use the default EW input parameters \cite{Haisch:2025jqr} that are compliant with the recent PDG review \cite{ParticleDataGroup:2024cfk}. The WC for the $Q_W$ operator (see Eq.~\eqref{eq:QW}) is initially set to 
$C_W/\Lambda^2 = 1\TeV^{-2}$ and then varied between $+1$ and $-1$.
The tree-level and one-loop SM and SMEFT amplitudes are computed with the \recola 2 \cite{Denner:2017wsf} library. The subtraction of QCD infrared singularities
is achieved in the FKS scheme \cite{Frixione:1995ms}. 
Both the factorisation and the renormalisation scale are set to the arithmetic average of the $\PW$- and $\PZ$-boson pole masses. The parton luminosities in the proton (PDFs), as well as the QCD coupling $\alpha_{\rm s}$, are evaluated through the {\sc LHAPDF} interface \cite{Buckley:2014ana}. The {\sc NNPDF31\_nlo\_as\_0118} PDF set \cite{Ball:2017nwa} is used.

For our phenomenological study we consider an inclusive setup,
\beq
\label{eq:incS}
81\GeV < M_{\mu^+\mu^-} < 101\GeV\,,
\eeq
and a boosted one,
\beq
\label{eq:booS}
81\GeV < M_{\mu^+\mu^-} < 101\GeV\,,\quad
\pt{\PZ}>200\GeV\,,\quad
\pt{\PW\PZ}<70\GeV\,.
\eeq
The two are simplified versions of ATLAS setups for polarisation studies \cite{ATLAS:2022oge,ATLAS:2024qbd}, obtained by
retaining only the selections affecting the boson kinematics and not the one of individual decay products. 
This allows to pursue the extraction of polarisation and spin-correlation coefficients in the same fashion as in Eqs.~\eqref{eq:VcoefPROJ} and \eqref{eq:VcoefPROJeft}. In the realistic case of an extraction from LHC data, an extrapolation of angular distributions 
from fiducial to uncut setups is required to undo the selections that are necessarily applied to decay products in any LHC event selection. 
This extrapolation is different for SM and beyond-the-SM underlying dynamics, owing to the
model-dependence of cut effects on polar- and azimuthal-angle distributions \cite{Grossi:2024jae}.

In the following we dub {SMEFT6} the predictions which include EFT effects up to dimension-six interference, \ie single-boson and di-boson angular coefficients are extracted as in Eq.~\eqref{eq:VcoefPROJeft6}. 
We dub {SMEFT8} all predictions including both interference and quadratic terms in the EFT expansion, with the coefficients extracted as in Eq.~\eqref{eq:VcoefPROJeft}.

We first present the $\alpha_{lm}^{(1)}$ angular coefficients, which fully determine the spin-density matrix for the $\PW$ boson.
The numerical results for coefficients that are numerically not compatible with zero are shown in Table~\ref{tab:nonvanish} for both considered setups.
\begin{table*}
\begin{center}
    \begin{tabular}{crrrrrr}
      \hline\rule{0ex}{2.7ex}
      \cellcolor{yellow!9} coeff.  
      & \cellcolor{yellow!9} LO
      & \cellcolor{yellow!9} NLO
      & \cellcolor{yellow!9} LO
      & \cellcolor{yellow!9} NLO
      & \cellcolor{yellow!9} LO
      & \cellcolor{yellow!9} NLO \\[0.1cm]
      \hline\\[-0.25cm]
      \cellcolor{yellow!9} & 
      \multicolumn{2}{c}{\cellcolor{yellow!9} SM } &
      \multicolumn{2}{c}{\cellcolor{yellow!9} SMEFT6 } &
      \multicolumn{2}{c}{\cellcolor{yellow!9} SMEFT8 } \\
      \hline\\[-0.25cm]
      \multicolumn{7}{l}{\cellcolor{blue!9} inclusive } \\
      \hline\\[-0.25cm]
 $\kappa^{(1)}_1\cdot\alpha^{(1)}_{ 10   }$ & $  -0.225 (1 ) $   & $  -0.234 (1 ) $ & $  -0.222 (1 ) $   & $  -0.234 (1 ) $&   $  -0.185 (1 ) $   & $  -0.207 (1 ) $\\[0.1cm]
 $\kappa^{(1)}_2\cdot\alpha^{(1)}_{ 20   }$ & $  0.384 (2 ) $   & $  0.320 (1 ) $ &$  0.371 (2 ) $   & $  0.308 (2 ) $ &  $  0.424 (2 ) $   & $  0.350 (1 ) $ \\[0.1cm]
 $\kappa^{(1)}_2\cdot\alpha^{(1)}_{ 2-2  }$ & $  -0.176 (2 ) $   & $  -0.137 (1 ) $ &$  -0.053 (2 ) $   & $  -0.064 (1 ) $ & $  -0.042 (1 ) $   & $  -0.056 (1 ) $\\[0.1cm]
\hline\\[-0.25cm]
      \multicolumn{7}{l}{\cellcolor{blue!9} boosted } \\
\hline\\[-0.25cm]
 $\kappa^{(1)}_1\cdot\alpha^{(1)}_{ 10   }$ & $  -0.228 (4 ) $   & $  -0.281 (7 ) $ & $  -0.201 (4 ) $   & $  -0.297 (7 ) $ &$  -0.017 (1 ) $   & $  -0.037 (1 ) $\\[0.1cm]
 $\kappa^{(1)}_2\cdot\alpha^{(1)}_{ 20   }$ & $  0.165 (8 ) $   & $  0.23 (1 ) $ & $  0.123 (8 ) $   & $  0.09 (1 ) $ & $  0.648 (2 ) $   & $  0.618 (3 ) $\\[0.1cm]
 $\kappa^{(1)}_2\cdot\alpha^{(1)}_{ 2-2  }$ & $ -0.005 (7 )  $   & $  -0.07 (1 ) $ &$  1.522 (8 ) $   & $  1.19 (1 ) $ & $  0.140 (1 ) $   & $  0.154 (2 ) $ \\[0.1cm]
    \hline
    \end{tabular}\qquad
  \end{center}
  \caption{ LO and NLO QCD polarisation coefficients for the $\PW$ boson in the inclusive and boosted setups described in Eq.~\eqref{eq:incS} and Eq.~\eqref{eq:booS}, respectively. SM, SMEFT6, and SMEFT8 results were obtained applying Eqs.~\eqref{eq:VcoefPROJ},~\eqref{eq:VcoefPROJeft6},~\eqref{eq:VcoefPROJeft}, respectively, and using $C_W/\Lambda^2=1\TeV^{-2}$ .
  Uncertainties in parentheses are statistical errors from numerical integration.
  \label{tab:nonvanish}
  }
\end{table*}
The NLO QCD corrections mostly affect the $l=2$ coefficients, while the changes in $\alpha^{(1)}_{10}$ are more moderate.
The QCD effects are much larger in the boosted setup rather than in the inclusive one, both in the SM and in the presence of SMEFT effects.
For $m=0$ coefficients, the inclusion of linear SMEFT terms has a minor impact compared to the quadratic EFT terms. This is somewhat expected as the quadratic term gives the leading change in the polarisation balance \cite{Haisch:2025jqr}, owing to 
the opening of helicity configurations that are suppressed in the SM \cite{Azatov:2016sqh,Helset:2017mlf}.
The situation is dramatically different for the only non-vanishing azimuthal coefficient $\alpha^{(1)}_{2-2}$. In the inclusive setup, its value is strongly diminished by the SMEFT linear terms, while much milder effects come from the quadratic terms.
In the boosted topology, this coefficients is extremely suppressed in the SM, while it gets positive owing to SMEFT contributions. While the SMEFT8 prediction is more or less of the same order of magnitude as in the inclusive setup, the
SMEFT6 prediction is anomalously large, signaling an ill-defined extraction of azimuthal coefficients when the SMEFT series is truncated at dimension six. This goes in the same direction as the analytical conclusions of Sect.~\ref{sec:analytic}.
Indeed, the LO SMEFT6 prediction leads to a trace of the squared $\rho_\PW$ matrix larger than 1 for $C_W/\Lambda^2=1\TeV^{-2}$, 
driven by the large value of $\alpha^{(1)}_{2-2}$. At NLO QCD the violation of the trace bound is achieved for larger values of the WC, owing to QCD corrections that sizeably diminish the purity of the system.
\begin{figure*}
    \centering
    \subfigure[Inclusive\label{fig:inclusive}]{\includegraphics[width=0.48\textwidth,page=1]{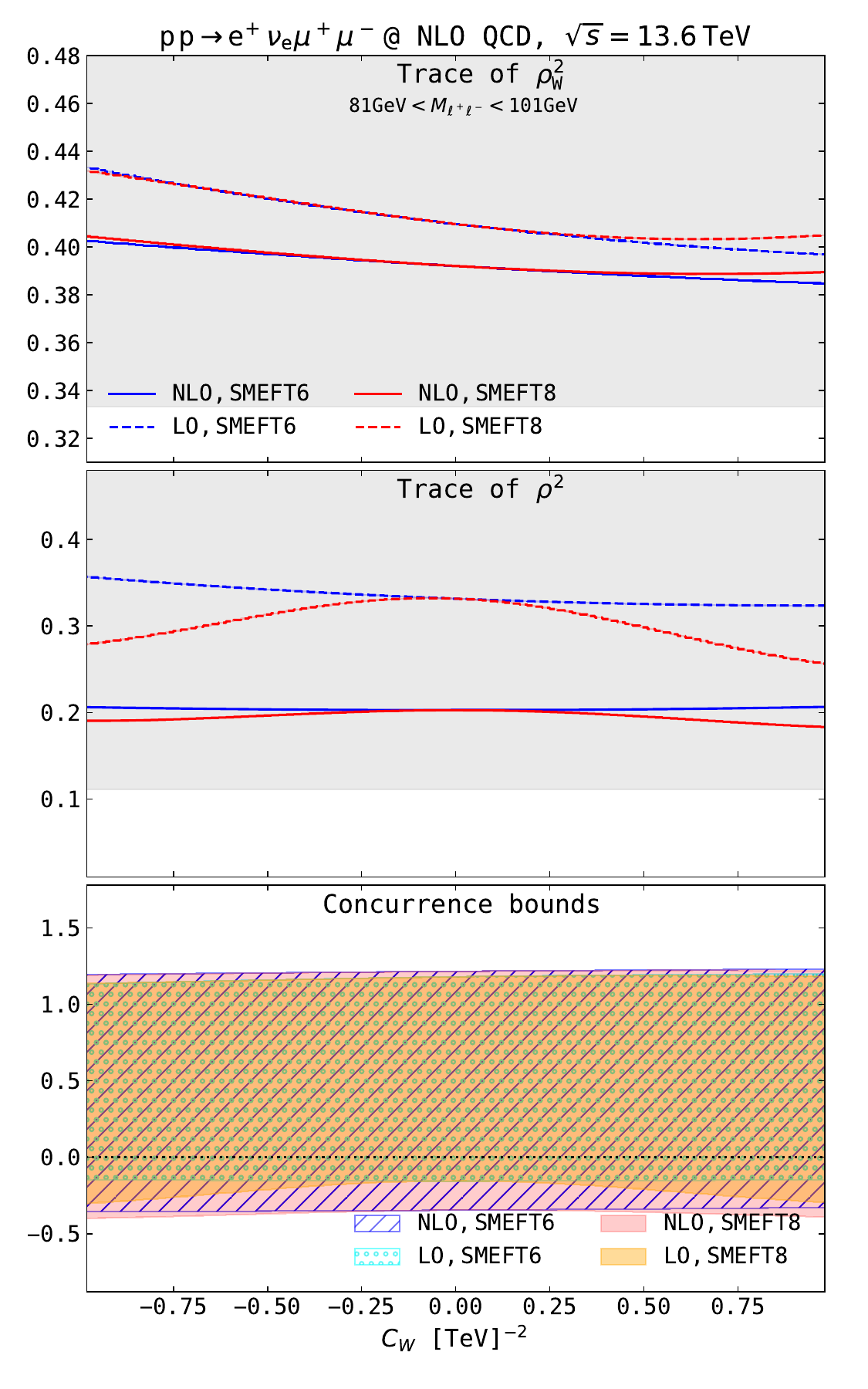}}
    \subfigure[Boosted\label{fig:boosted}]{\includegraphics[width=0.48\textwidth,page=1]{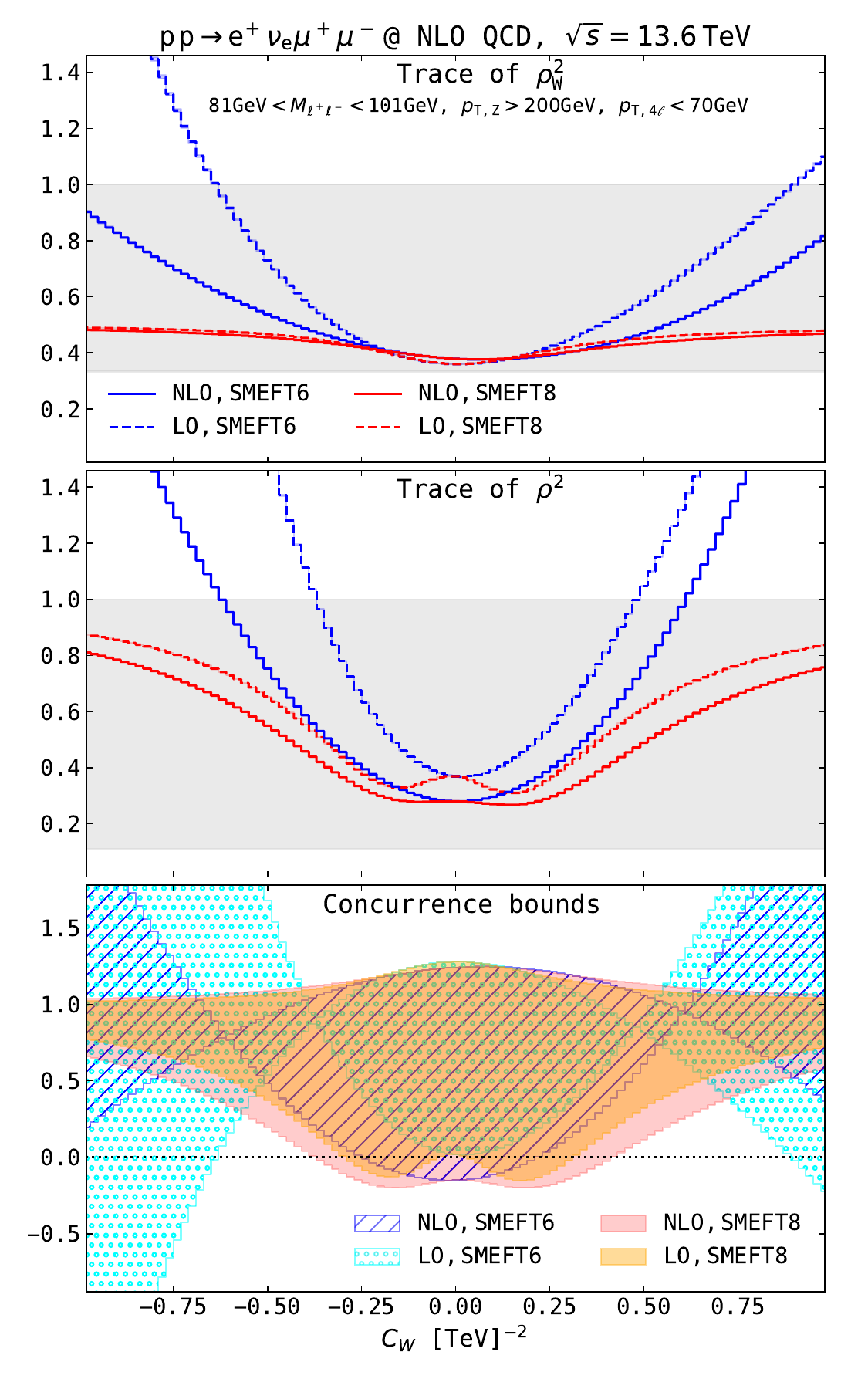}}
    \caption{
    Dependence of purity and concurrence markers on the Wilson coefficient $C_W$ for inclusive $\PW\PZ$ production at the LHC:
    the trace of the squared single-boson ($\PW$) spin-density matrix (upper panels), 
    the trace of the squared $\rho$ matrix associated to the overall $\PW\PZ$ system (middle panels), and
    the concurrence lower and upper bounds for the $\PW\PZ$ system (lower panels).
    Results have been computed at both LO and NLO QCD accuracy, including SMEFT effects up to linear (SMEFT6) and quadratic (SMEFT8) level.
    The inclusive (boosted) setup described in Eq.~\eqref{eq:incS} (Eq.~\eqref{eq:booS}) is considered in the left (right) plot.
    In the upper and middle panels, the gray bands correspond to allowed values for generic hermitian, semi-positive definite, unit-trace spin-density matrices.
    }
    \label{fig:obs}
\end{figure*}
This can be observed in the top panels of Fig.~\ref{fig:obs}. While in the inclusive setup the EFT effects are milder and not sufficient to spot the ill definition of the SMEFT6 predictions, in the boosted setup the sizeable EFT contributions
expose in a striking manner the issues of 
Eq.~\ref{eq:rhoSMEFT6}. In kinematic topologies where the spin state has maximal purity, this effect would be even larger.
From the top panel of Fig.~\ref{fig:boosted} it is crystal clear that including quadratic terms is of paramount importance for a proper definition of the $\PW$-boson spin-density matrix, regularising the ill-defined behaviour of a $\rho_\PW$
matrix as in Eq.~\ref{eq:rhoSMEFT6}.
It is also important to notice that for large values of the WC and for boosted topologies the inclusion of NLO QCD corrections has a lower impact on SMEFT8 predictions than on SMEFT6 ones. 
As a last comment on sigle-boson coefficients, one could naively argue that the SMEFT6 predictions in the boosted topology allow to constrain better the $C_W$ value rather than using SMEFT8 ones. While tempting from the EFT-fit viewpoint, we refrain from drawing conclusions in this direction, not only because of the ill-defined spin-density matrix, but also also because in this kinematic regime the SMEFT predictions start violating perturbative unitarity, and therefore one can expect that observables like the transverse momentum of the boson are expected to constrain better the WCs compared to angular coefficients. 

Before discussing two-boson correlations, we point out that the results for the $\PZ$-boson spin-density matrix are very similar to those for the $\PW$ boson in Table~\ref{tab:nonvanish} and upper panels of Fig.~\ref{fig:obs}, therefore we have not shown them explicitly in the text. 

Looking at the red curves in the middle panels in Fig.~\ref{fig:obs}, we observe that for a large WC value in the boosted topology the purity of the $\PW\PZ$-system spin state ($\approx 0.8$) is higher than the one of the $\PW$ boson ($\approx 0.5$).
Since the helicity rules select $\PW_\pm\PZ\pm$ configurations (with the same share of left and right polarisations) and the size of single-boson azimuthal coefficients is moderate, the enhancement of the di-boson purity only comes from di-boson correlations and in particular from large values of 
the polar coefficient $\gamma_{1010}$, which depends on the relative share of left and right helicity of the two bosons \cite{Grossi:2024jae}, and of the azimuthal correlations $\gamma_{2\pm22\mp2}$.
On the contrary, in the inclusive setup the purity of the $\PW\PZ$ system ($\approx 0.2$) is lower than the one of the $\PW$ ($\approx 0.4$), owing to very small azimuthal correlations.
Overall, the inclusion of QCD corrections diminishes the purity of both the $\PW$ and the $\PW\PZ$ systems, as expected especially from the large real-radiation corrections that drive the NLO QCD correction even in the presence of 
contraints on the hadronic activity ($\pt{\PW\PZ}<70\GeV$).
As discussed for $\rho_\PW$, the SMEFT6 description violates the constraint ${\rm Tr}[\rho^2]\leq 1$ as well in the boosted regime. This violation appears for even smaller values of the $C_W/\Lambda^2$ than in the case of $\rho_\PW$, for both LO and NLO QCD predictions. 

We have discussed at the end of Sect.~\ref{sec:analytic} that the traces of the squared single- and di-boson spin-density matrices can be combined into a lower and an upper bounds for the concurrence marker for spin entanglement.
The corresponding numerical results are presented in the lower panels of Fig.~\ref{fig:obs}.
Because of the violation of the ${\rm Tr}[\rho^2]\leq 1$ and ${\rm Tr}[\rho_V^2]\leq 1$ constraints in the SMEFT6 picture, the upper bound on the concurrence crosses the lower bound for absolute values of the $Q_W$-operator WC below $0.5$. Even more strikingly than for the di-boson spin purity, the SMEFT6 description of the spin density matrix fails in the boosted regime and for values of the WC that are still allowed by EFT fits.
The inclusion of the quadratic terms in the SMEFT8 predictions regularises the results.
In the inclusive setup the SMEFT6 and SMEFT8 bounds look substantially the same and do not allow for any conclusion about spin entanglement, as the lower bound is negative. On the contrary, the SMEFT8 predictions in the boosted setup show a
lower concurrence bound that exceeds zero for $|C_W/\Lambda^2|\approx 0.4(0.3)$ at NLO QCD (LO), highlighting a sizeable entanglement of the $\PW\PZ$ spin state. As for the purity markers, the NLO QCD corrections diminish the lower bound of concurrence, pointing towards quantum-decoherence effects from QCD radiation. 
We note that at LO in the boosted setup, while for values of the WC around 0.2 the lower bound is negative, in the strict SM limit it becomes slightly positive, in principle allowing for a claim of entanglement. However, introducing NLO QCD corrections makes the lower limit negative also for the SM. This means that, in spite of the possibility to optimise phase-space selections to enhance entanglement \cite{Fabbrichesi:2023cev,Aoude:2023hxv}, the incusion of higher-order QCD corrections could sizeably change the LO pictures, even in the presence of realistic jet vetoes. 

Although beyond the purposes of this work, we believe that the inclusion of spin correlations and entanglement markers as inputs to EFT fits is worth a try, even if it is not clear that their inclusion would improve the WC bounds. Even more importantly, the practical utility of QI-inspired observables in SMEFT fits would only become clear after extending our study to multiple SMEFT operators. This should be done with care to avoid similar conceptual issues as those we discussed for SMEFT6 predictions in this work. For example, including two independent dimension-six operators in the definition of the spin-density matrix would require to include not only quadratic terms for both operators, but also the interference between them. 

\section{Conclusion}\label{sec:conclusion}
We have presented a comprehensive analysis of spin correlations in the inclusive LHC production of $\PW\PZ$ pairs in the fully leptonic channel. We have included NLO corrections in the strong coupling and new-physics contributions introduced by a SMEFT dimension-six operator that leads to anomalous electroweak TGCs.
We have carried out the full quantum-state tomography of the considered di-boson system,
and connected the resulting entries of the spin-density matrix to standard markers of spin purity and entanglement.

We have shown that NLO QCD effects are very large on many polarisation and spin-correlation coefficients, both in the SM and in the SMEFT. Furthermore, the inclusion of higher-order QCD corrections was proved to diminish the level of purity and of  entanglement of the two-boson spin state, signaling decoherence effects.

The insertion of CP-even SMEFT vertices modifying the TGCs has a moderate impact on angular coefficients in inclusive setups, while larger effects and non-trivial features appear for bosons with large transverse momenta.

Strikingly, in boosted topologies and for certain ranges of the considered SMEFT WC, 
the trace of the squared spin-density matrix exceeds one if the angular coefficients are extracted truncating the SMEFT expansion at dimension-six, \ie no quadratic EFT terms. Consequently, the naive interpretation of angular coefficients in terms of the spin-density-matrix entries becomes cumbersome. 
A meaningful spin interpretation for arbitrary values of WCs and arbitrary kinematic regimes can be obtained only if quadratic EFT terms are properly included in the definition of the spin-density matrix and in the calculation of angular coefficients. This represents the main result of our study.

We also conclude that in realistic kinematic regimes, even boosted ones employed for recent LHC polarisation measurements, as well as for realistic values of the SMEFT WC allowed by global fits, the concurrence bounds do not allow to show that the WZ system is spin-entangled. This is especially true after including NLO QCD corrections, while the LO results look more promising in this sense.
Once again, we stress that the LO picture of spin correlations and entanglement markers is limited as far as the LHC production of boson pairs is concerned, making it vital to include at least NLO (and possibly NNLO, which is now within reach) QCD corrections.

This study was focused on a single SMEFT operator modifying the TGCs for reasons of simplicity. However, we believe our conclusions would apply also to other operators, possibly affecting both the production and the decay of EW bosons.

\section*{Acknowledgements}
We would like to thank Jakob Linder and Fabio Maltoni for useful discussions.
The authors acknowledge support from the COMETA EU COST Action (CA22130).
GP is funded by the EU Horizon Europe research and innovation programme under the Marie-Sk\l{}odowska Curie Action (MSCA) ``POEBLITA - POlarised Electroweak Bosons at the LHC with Improved Theoretical Accuracy'', grant agreement Nr.~101149251 (CUP H45E2300129000). ER is supported by the Italian Ministry of University and Research (MUR), with EU funds (NextGenerationEU), through the PRIN2022 grant agreement Nr.~20229KEFAM (CUP H53D23000980006). 

%%%%%%%%%%%%%%%%%%%%%%%%%%%%%%%%%

\bibliographystyle{JHEPmod}
\bibliography{polvv}

\end{document}